\documentclass[11pt]{article}
\usepackage{latexsym}
\usepackage{amssymb}
\usepackage{graphics,graphicx}
\usepackage{tabularx}
\usepackage{amsfonts}
\usepackage{amsmath}
\usepackage{enumerate}
\usepackage{booktabs}

\newcommand{\footmsg}[1]{%
  \let\temp\thempfn%
  \def\thempfs{}
  \footnotetext{#1}
  \let\tempfn\temp}

\begin{document}

%Definitions: general
\newcommand{\singlespace} {\baselineskip=12pt
\lineskiplimit=0pt \lineskip=0pt }
\def\ds{\displaystyle}

%Definitions: equations
\newcommand{\beq}{\begin{equation}}
\newcommand{\eeq}{\end{equation}}
\newcommand{\lb}{\label}
\newcommand{\beqar}{\begin{eqnarray}}
\newcommand{\eeqar}{\end{eqnarray}}
\newcommand{\barr}{\begin{array}}
\newcommand{\earr}{\end{array}}

\newcommand{\jump}{\parallel}

\def\c{{\circ}}

\newcommand{\Ehat}{\hat{E}}
\newcommand{\That}{\hat{\bf T}}
\newcommand{\Ahat}{\hat{A}}
\newcommand{\chat}{\hat{c}}
\newcommand{\shat}{\hat{s}}
\newcommand{\khat}{\hat{k}}
\newcommand{\muhat}{\hat{\mu}}
\newcommand{\mc}{M^{\scriptscriptstyle C}}
\newcommand{\mei}{M^{\scriptscriptstyle M,EI}}
\newcommand{\mec}{M^{\scriptscriptstyle M,EC}}

\newcommand{\hbeta}{{\hat{\beta}}}
\newcommand{\rec}[2]{\left( #1 #2 \ds{\frac{1}{#1}}\right)}
\newcommand{\rep}[2]{\left( {#1}^2 #2 \ds{\frac{1}{{#1}^2}}\right)}
\newcommand{\derp}[2]{\ds{\frac {\partial #1}{\partial #2}}}
\newcommand{\derpn}[3]{\ds{\frac {\partial^{#3}#1}{\partial #2^{#3}}}}
\newcommand{\dert}[2]{\ds{\frac {d #1}{d #2}}}
\newcommand{\dertn}[3]{\ds{\frac {d^{#3} #1}{d #2^{#3}}}}

\def\bob{{\, \underline{\overline{\otimes}} \,}}

\def\ob{{\, \underline{\otimes} \,}}
\def\scalp{\mbox{\boldmath$\, \cdot \, $}}
\def\gdp{\makebox{\raisebox{-.215ex}{$\Box$}\hspace{-.778em}$\times$}}

\def\daa{\makebox{\raisebox{-.050ex}{$-$}\hspace{-.550em}$: ~$}}

\def\mK{\mbox{${\mathcal{K}}$}}
\def\cK{\mbox{${\mathbb {K}}$}}

%Definitions: integrals
\def\Xint#1{\mathchoice
   {\XXint\displaystyle\textstyle{#1}}%
   {\XXint\textstyle\scriptstyle{#1}}%
   {\XXint\scriptstyle\scriptscriptstyle{#1}}%
   {\XXint\scriptscriptstyle\scriptscriptstyle{#1}}%
   \!\int}
\def\XXint#1#2#3{{\setbox0=\hbox{$#1{#2#3}{\int}$}
     \vcenter{\hbox{$#2#3$}}\kern-.5\wd0}}
\def\ddashint{\Xint=}
\def\fpint{\Xint=}
\def\dashint{\Xint-}
\def\cpvint{\Xint-}
\def\intl{\int\limits}
\def\cpvintl{\cpvint\limits}
\def\fpintl{\fpint\limits}
\def\ointl{\oint\limits}

\def\half{{\scriptstyle{\frac{1}{2}}}}

\def\bA{{\bf A}}
\def\ba{{\bf a}}
\def\bB{{\bf B}}
\def\bb{{\bf b}}
\def\bc{{\bf c}}
\def\bC{{\bf C}}
\def\bD{{\bf D}}
\def\bE{{\bf E}}
\def\be{{\bf e}}
\def\bbf{{\bf f}}
\def\bF{{\bf F}}
\def\bG{{\bf G}}
\def\bg{{\bf g}}
\def\bi{{\bf i}}
\def\bH{{\bf H}}
\def\bK{{\bf K}}
\def\bL{{\bf L}}
\def\bM{{\bf M}}
\def\bN{{\bf N}}
\def\bn{{\bf n}}
\def\bm{{\bf m}}
\def\b0{{\bf 0}}
\def\bo{{\bf o}}
\def\bX{{\bf X}}
\def\bx{{\bf x}}
\def\bP{{\bf P}}
\def\bp{{\bf p}}
\def\bQ{{\bf Q}}
\def\bq{{\bf q}}
\def\bR{{\bf R}}
\def\bS{{\bf S}}
\def\bs{{\bf s}}
\def\bT{{\bf T}}
\def\bt{{\bf t}}
\def\bU{{\bf U}}
\def\bu{{\bf u}}
\def\bv{{\bf v}}
\def\bw{{\bf w}}
\def\bW{{\bf W}}
\def\by{{\bf y}}
\def\bz{{\bf z}}

\def\T{{\bf T}}
\def\Te{\textrm{T}}

\def\e{{\rm{e}}}
\def\Id{{\bf I}}
\def\p{{\rm{p}}}
\def\t{{\rm{t}}}
\def\bxi{\mbox{\boldmath${\xi}$}}
\def\balpha{\mbox{\boldmath${\alpha}$}}
\def\bbeta{\mbox{\boldmath${\beta}$}}
\def\bepsilon{\mbox{\boldmath${\epsilon}$}}
\def\bvarepsilon{\mbox{\boldmath${\varepsilon}$}}
\def\bomega{\mbox{\boldmath${\omega}$}}
\def\bphi{\mbox{\boldmath${\phi}$}}
\def\bsigma{\mbox{\boldmath${\sigma}$}}
\def\bfeta{\mbox{\boldmath${\eta}$}}
\def\bDelta{\mbox{\boldmath${\Delta}$}}
\def\btau{\mbox{\boldmath $\tau$}}
\def\bSS{\mbox{\boldmath $S$}}
\def\bII{\mbox{\boldmath $I$}}

\def\tr{{\rm tr}}
\def\dev{{\rm dev}}
\def\div{{\rm div}}
\def\Div{{\rm Div}}
\def\Grad{{\rm Grad}}
\def\grad{{\rm grad}}
\def\Lin{{\rm Lin}}
\def\Sym{{\rm Sym}}
\def\Skw{{\rm Skew}}
\def\abs{{\rm abs}}
\def\Re{{\rm Re}}
\def\Im{{\rm Im}}
\def\sign{{\rm sign}}

\def\capB{\mbox{\boldmath${\mathsf B}$}}
\def\capC{\mbox{\boldmath${\mathsf C}$}}
\def\capD{\mbox{\boldmath${\mathsf D}$}}
\def\capE{\mbox{\boldmath${\mathsf E}$}}
\def\capG{\mbox{\boldmath${\mathsf G}$}}
\def\tcapG{\tilde{\capG}}
\def\capH{\mbox{\boldmath${\mathsf H}$}}
\def\capK{\mbox{\boldmath${\mathsf K}$}}
\def\capL{\mbox{\boldmath${\mathsf L}$}}
\def\capM{\mbox{\boldmath${\mathsf M}$}}
\def\capR{\mbox{\boldmath${\mathsf R}$}}
\def\capW{\mbox{\boldmath${\mathsf W}$}}

%imaginary unit
\def\i{\mbox{${\mathrm i}$}}

\def\mC{\mbox{\boldmath${\mathcal C}$}}

\def\mB{\mbox{${\mathcal B}$}}
\def\mE{\mbox{${\mathcal{E}}$}}
\def\mL{\mbox{${\mathcal{L}}$}}
\def\mK{\mbox{${\mathcal{K}}$}}
\def\mV{\mbox{${\mathcal{V}}$}}

\def\C{\mbox{\boldmath${\mathcal C}$}}
\def\E{\mbox{\boldmath${\mathcal E}$}}

%Definitions: journals
\def\ACME{{ Arch. Comput. Meth. Engng.\ }}
\def\ARMA{{ Arch. Rat. Mech. Analysis\ }}
\def\AMR{{ Appl. Mech. Rev.\ }}
\def\ASCEEM{{ ASCE J. Eng. Mech.\ }}
\def\acta{{ Acta Mater. \ }}
\def\ACTA{{ Acta Mechanica \ }}
\def\AMM{{ Acta Metall. Mater. \ }}
\def\CMAME {{ Comput. Meth. Appl. Mech. Engrg.\ }}
\def\CRAS{{ C. R. Acad. Sci., Paris\ }}
\def\EFM{{ Eng. Fracture Mechanics\ }}
\def\EJMA{{ Eur.~J.~Mechanics-A/Solids\ }}
\def\IJES{{ Int. J. Eng. Sci.\ }}
\def\IJF{{\it Int. J. Fracture\ }}
\def\IJMS{{ Int. J. Mech. Sci.\ }}
\def\IJNAMG{{ Int. J. Numer. Anal. Meth. Geomech.\ }}
\def\IJP{{ Int. J. Plasticity\ }}
\def\IJSS{{ Int. J. Solids Structures\ }}
\def\IngA{{ Ing. Archiv\ }}
\def\JAM{{ J. Appl. Mech.\ }}
\def\JAP{{ J. Appl. Phys.\ }}
\def\JEM{{J. Engrg. Mech., ASCE\ }}
\def\JE{{ J. Elasticity\ }}
\def\JM{{ J. de M\'ecanique\ }}
\def\JMPS{{ J. Mech. Phys. Solids\ }}
\def\Macro{{ Macromolecules\ }}
\def\MOM{{ Mech. Materials\ }}
\def\MMS{{ Math. Mech. Solids\ }}
\def\MMT{{\it Metall. Mater. Trans. A}}
\def\MPCPS{{ Math. Proc. Camb. Phil. Soc.\ }}
\def\MSE{{ Mater. Sci. Eng.}}
\def\PMPS{{ Proc. Math. Phys. Soc.\ }}
\def\PRE{{ Phys. Rev. E\ }}
\def\PRSL{{ Proc. R. Soc.\ }}
\def\rock{{ Rock Mech. and Rock Eng.\ }}
\def\QAM{{ Quart. Appl. Math.\ }}
\def\QJMAM{{ Quart. J. Mech. Appl. Math.\ }}
\def\SCRMAT{{ Scripta Mater.\ }}
\def\SM{{\it Scripta Metall. }}
\def\JACS{{ J. Am. Ceram. Soc.\ }}
\def\JMS{{ J. Materials Sci.\ }}
\def\PT{{ Powder Tech.\ }}

% segue comando pazzesco di zaccaria

\def\salto#1#2{
%\left[\mbox{\hspace{-#1em}}\left[#2\right]\mbox{\hspace{-#1em}}\right]}
\left[\mbox{\hspace{-#1em}}\left[#2\right]\mbox{\hspace{-#1em}}\right]}

\def\medio#1#2{
%\left[\mbox{\hspace{-#1em}}\left[#2\right]\mbox{\hspace{-#1em}}\right]}
\mbox{\hspace{-#1em}}<#2>\mbox{\hspace{-#1em}}}

%\def\salto{
%\left[\mbox{\hspace{-#1em}}\left[#2\right]\mbox{\hspace{-#1em}}\right]}
%[[#1]]}

%\newcommand{\salto}[1]{\ds{\|#1\|}}

%dopodiche' scrivi il comando
%\salto{*}{**}
%dove * e' un numero mentre ** e' quello che tu vuoi tra parentesi.
%Il numero serve a far si che le due parentesi quadre che comporranno
%l'unica parentesi che tu vuoi siano ben posizionate. Tale numero varia a
%seconda di chi sia **.

\title {Experimental investigation of the elastoplastic response of aluminum silicate spray dried powder during cold compaction}

\author{ F. Bosi$^{a,b}$,  A. Piccolroaz$^a$, M. Gei$^{a,b}$,
F. Dal Corso$^a$, A. Cocquio$^b$ and D. Bigoni$^{a,}$\footnote{Corresponding author. Tel.: +39 0461282507; {\it E-mail address}: bigoni@ing.unitn.it (D. Bigoni)}\\
\\
\small{$^a$ Department of Civil, Environmental and Mechanical Engineering,}\\
\small{University of Trento, via Mesiano 77, I-38123 Trento, Italy}\\
\small{$^b$ R\&D Department, SACMI Imola s.c.,}\\
\small{Via Provinciale Selice 17/A, I-40026 Imola (BO), Italy}\\
}

\date{}
\maketitle

\begin{abstract}

Mechanical experiments have been designed and performed to investigate the elasto-plastic behaviour
of green bodies formed from an aluminum silicate spray dried powder used for tiles production.
Experiments have been executed on samples obtained from cold compaction into a cylindrical mould and include:
uniaxial strain, equi-biaxial flexure and high-pressure triaxial compression/extension tests.
Two types of powders have been used to realize the green body samples, differing in the values of water content,
which have been taken equal to those usually employed in the industrial forming of traditional ceramics.
Yielding of the green body during compaction has been characterized in terms of yield surface shape, failure envelope, and
evolution of cohesion and void ratio with the forming pressure, confirming the validity of previously proposed
constitutive models for dense materials obtained through cold compaction of granulates.

\end{abstract}


\section{Introduction}

Compaction of aluminum silicate powder is the key process to produce tiles and traditional ceramics. This process
is strictly connected to the plastic properties of the granulate and of the green bodies obtained at different forming pressures,
so that the investigation of their mechanical behavior paves the way to the constitutive modelling of the material in its different compaction states.
The knowledge of the constitutive features of green bodies allows the simulation of the forming process of ceramics and thus its optimization. Since
the industrial production of traditional ceramics is currently highly energy expensive, optimization of this process becomes imperative in the perspective
of sustainable development. It may therefore seem surprising that the experimental investigation of the plastic behaviour
of green bodies has received little attention so far (Baklouti et al., 1997a; b; 
Carneim and Green, 2001; Kim et al. 2000; 2003; Lee and Kim, 2008; Maity and Sarkar, 1996; 
\"Ozkan and Briscoe, 1997; Zeuch et al., 2001; Zhang and Green, 2002), particularly for traditional ceramics.

In this article, the mechanical characterization of the elasto-plastic behaviour of green bodies
(realized  through cold compaction of aluminum silicate powder)
is addressed to the four most important features, namely, the investigation of: (i.)
the elastic domain as delimited by the yield locus; (ii.) the failure envelope (or critical line); and the laws of increase of (iii.) density and (iv.) of cohesion, both with the forming pressure. The performed experiments include uniaxial strain of granulates, equi-biaxial flexure and high-pressure
triaxial compression/extension tests on green bodies.
The yield locus is shown to be representable with the yield surface proposed by Bigoni and Piccolroaz (2004; see also Bigoni, 2012) and 
the failure envelope with a curved Coulomb criterion, while the variations of density
and of cohesion with the forming pressure confirm the laws of Cooper and Eaton (1962) and Piccolroaz et al. (2006 a), respectively.
It can be finally pointed out that, in addition to support the suitability of the elastoplastic coupled model for ceramic powder compaction
proposed by Piccolroaz et al. (2006 a; b), the experimental data obtained in the present article are fundamental to perform the material parameter
calibration necessary to simulate the forming process of traditional ceramics (Penasa et al., 2013; Stupkiewicz et al., 2013).
%%%%%%%%%%%%%%%%%%%%%%%%%%%%%%%%%%%%%%%%%%%%%%%%%%%%%%%%%%%%%%%%%%%%%%%%%%%%%%%%%%%%%

\section{Forming and evolution of void ratio for aluminum silicate green bodies}\lb{esp}

\subsection{Tested material}

The considered material is the aluminum silicate spray dried powder, labeled I14730, shown in Fig. \ref{micrografia} and tested at two different water contents, namely, $w=5.5\%$ and $w=7.5\%$,
corresponding to  values used in the industrial forming of traditional ceramics.
The powders have been manufactured by Sacmi s.c. (Imola, Italy) and have the granulometric
properties reported in Table \ref{granulom}.
%%%%%%%%%%%%%%%%%%%%%%%%%%%%%%%%%%%%%%%%%%%%%%%%%%%%%%%%%%%%%%%%%%%%%%
\begin{figure}[!htcb]
  \begin{center}
      \includegraphics[width= 15 cm]{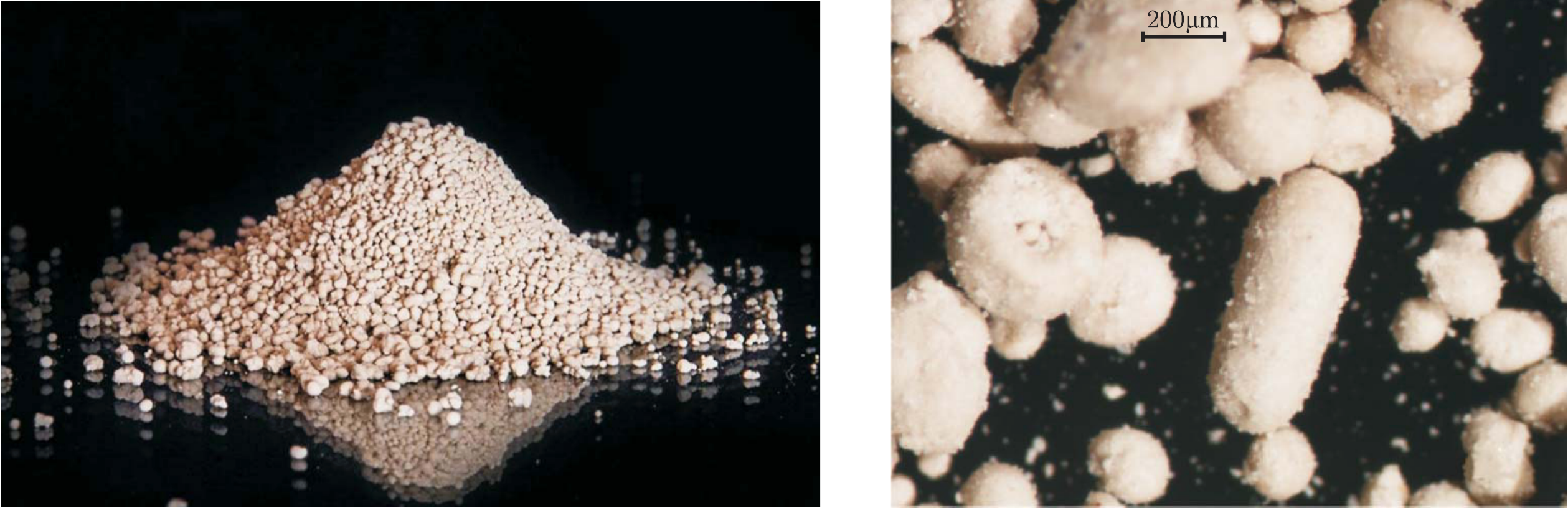}
\caption{\footnotesize I14730 aluminum silicate spray dried powder in the loose state: bulk material (left)
and micrograph at 31.5x (right, obtained with a Nikon
stereoscopic microscope Nikon SMZ-800 equipped with Nikon Plan Apo 0.5x objective and a Nikon DS-Fi1
high-definition color camera head). The granules have an irregular surface and a mean diameter of 425 $\mu$m.}
\lb{micrografia}
  \end{center}
\end{figure}
%%%%%%%%%%%%%%%%%%%%%%%%%%%%%%%%%%%%%%%%%%%%%%%%%%%%%%%%%%%%%%%%%%%%%%
%%%%%%%%%%%%%%%%%%%%%%%%%%%%%%%%%%%%%%%%%%%%%%%%%%%%%%%%%%%%%%%%%%%%%%
\begin{table} \caption{\footnotesize{Granulometric properties of the tested aluminum silicate spray dried powder}} \label{granulom}
\begin{center}
\begin{tabular}{cc}
\toprule
\textbf{sieve residue}  & \textbf{percentage} \\
\midrule
$> 600\,\mu$m & 26.5\% \\
$> 425\,\mu$m & 43.0\% \\
$> 300\,\mu$m & 15.5\% \\
$> 250\,\mu$m & 7.3\% \\
$> 180\,\mu$m & 4.6\% \\
$< 125\,\mu$m & 3.1\% \\
\bottomrule
\end{tabular}
\end{center}
\end{table}

%%%%%%%%%%%%%%%%%%%%%%%%%%%%%%%%%%%%%%%%%%%%%%%%%%%%%%%%%%%%%%%%%%%%%%

In order to obtain the particle (or true) density  $\rho_t$ of the powder, similarly to Della Volpe et al. (2007), an
helium pycnometer (AccuPyc 1330TC by Micromeritics) has been used after drying the powder (through heating at 100 $^\circ$C for one hour)
to evaluate its effective volume.
The mean value of the particle density was found to be $\rho_t= 2.599\;\textup{g}/\textup{cm}^3\;\pm 0.026 \%$.

\subsection{Forming through uniaxial strain and void ratio evolution}\label{ciccia}

Green body tablets have been realized through uniaxial strain forming into a (30 mm diameter) mould,
filled until a height of 4 mm with the aluminum silicate powder. Compaction has been imposed under
displacement control of the mould's punch (at a speed of 1.5 mm/min),
until final axial compression stresses\footnote{The convention of positive stress when compressive is adopted throughout this article.
}
of $\sigma_1=\left\{5,\,10,\, 30,\,40,\,60,\,80\right\}$ MPa has been reached and
the tablet finally extracted from the mould after unloading.
Five tablets for each final forming pressure have been produced.

The above-described uniaxial compaction tests have provided load $F$ vs. displacement $u_1$ curves,
which have been corrected considering the deformation
of the system (load cell, mould, movable crossbeam).\footnote{The deformability of the loading system
has been measured by performing a test on a steel sample.}

In order to obtain the variations in the density $\rho$ and in the void index $\Delta e$ as functions of the forming pressure $\sigma_1$ reported in Fig. \ref{densvuoti}, the load-displacement curves obtained from uniaxial strain have been elaborated through the equations 
\beq
\rho=\frac{m}{A_0\,(h_0-u_1)}, \qquad \sigma_1=\frac{F}{A_0}, \qquad \Delta e=-\frac{\rho_t\,A_0\,u_1}{m\,(1-w)},
\eeq
where $m$ is the mass of the powder inserted in the mould, $A_0$ is the transversal cross section area (706.86 mm$^2$),
$h_0$ is the initial height of the tablet  (4 mm) and $w$ is the water content. 
%%%%%%%%%%%%%%%%%%%%%%%%%%%%%%%%%%%%%%%%%%%%%%%%%%%%%%%%%%%%%%%%%%%%%%
\begin{figure}[!htcb]
  \begin{center}
      \includegraphics[width= 15 cm]{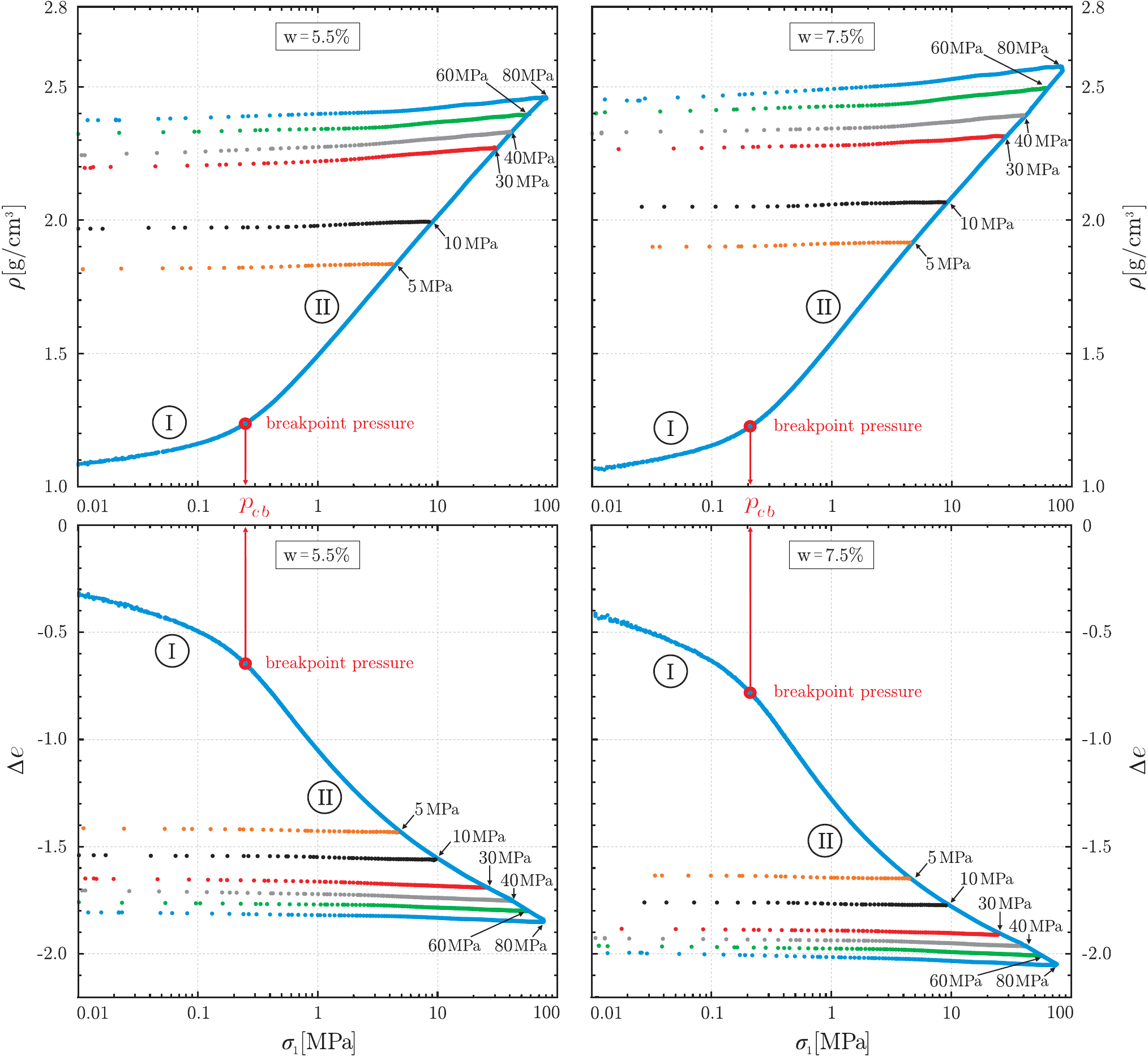}
\caption{\footnotesize Logarithmic plots of density $\rho$ (upper part) and variation in the
void index $\Delta e$ (lower part) vs. forming uniaxial pressure  $\sigma_1$ for water contents of $w=5.5\%$ (left) and $w=7.5\%$ (right).
Different forming pressures are considered, $\sigma_1=\left\{5,\,10,\, 30,\,40,\,60,\,80\right\}$ MPa,
after which the tablet has been eventually unloaded.
Note that two different phases of the compaction process can be identified within the loading curves,
the change in inclination of the curve is marked by \textit{breakpoint pressure}. The mean value of five tests with the same set-up
(water content and forming pressure) is reported.}
\lb{densvuoti}
  \end{center}
\end{figure}
%%%%%%%%%%%%%%%%%%%%%%%%%%%%%%%%%%%%%%%%%%%%%%%%%%%%%%%%%%%%%%%%%%%%%%
In Fig. \ref{densvuoti}, where the curves are mean values obtained through averaging results on five samples for each forming pressure, the first two of the three phases of the compaction process of
ceramic powders can be recognized: (I) granule sliding and rearrangement and  (II) granule deformation;
the change of phase is marked by the so-called \lq breakpoint pressure' $p_{cb}$.

As explained by Piccoloraz et al.  (2006 a), during elastic deformation of a powder still in the granular state, the increment in the void ratio $\Delta e^e$ depends on the current mean pressure $p=\left(\sigma_1+2\sigma_3\right)/3$ as
\beq
\Delta e^e=-\kappa\, \textup{log}\frac{p}{p_0},
\eeq
where the (dimensionless) logarithmic elastic bulk modulus $\kappa$ of the powder can be evaluated
 through a linear regression of the data in the elastic phase I (as the mean value of the results obtained from forming at different pressures)
at different water content. In particular, the values $\kappa=0.080 \pm0.003\%$ and 
$\kappa= 0.099 \pm0.003\%$ have been found for w=5.5\% and w = 7.5\%, respectively.

Finally, the density of the tablets, formed at different uniaxial  pressures, has been
measured with a mercury pycnometer, Fig. \ref{edom}, and has been found to be about 90$\%$ of the values observed
at the unloaded stage of the green body inside the mould, Fig.  \ref{densvuoti}, due to the spring-back.
%%%%%%%%%%%%%%%%%%%%%%%%%%%%%%%%%%%%%%%%%%%%%%%%%%%%%%%%%%%%%%%%%%%%%%
\begin{figure}[!htcb]
  \begin{center}
      \includegraphics[width= 8 cm]{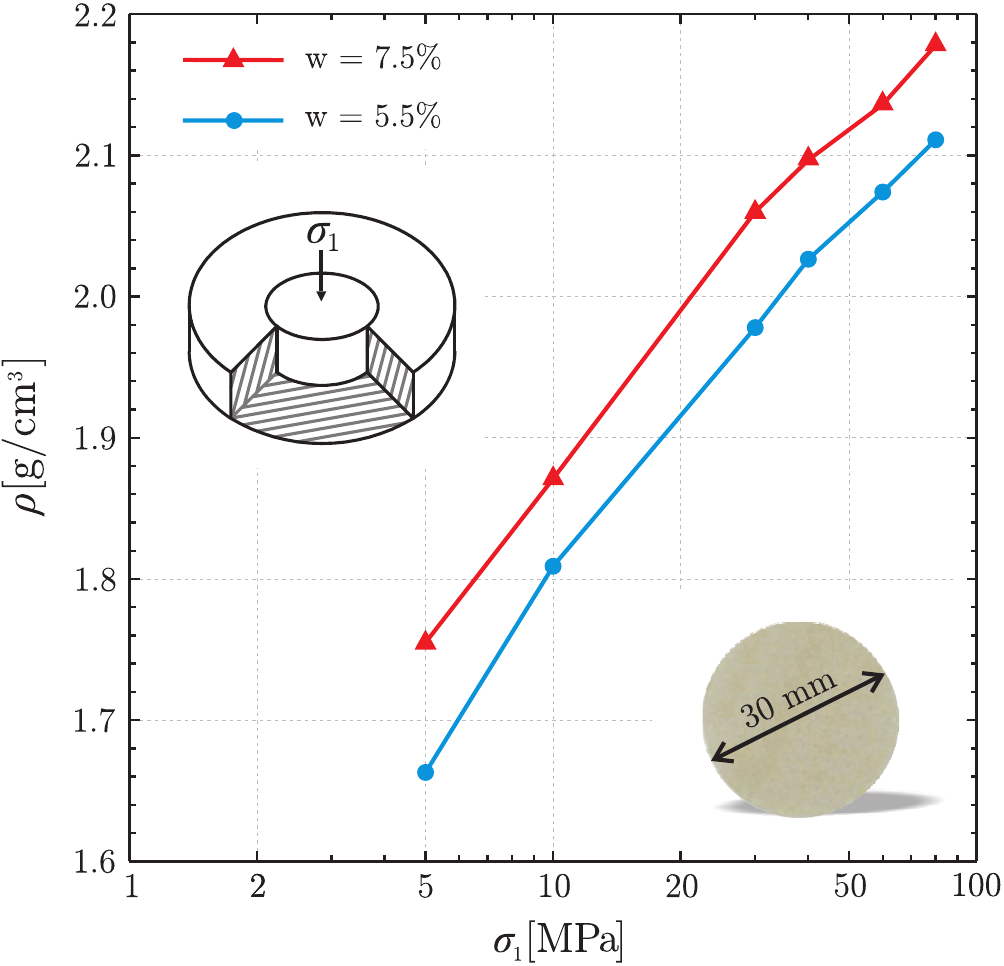}
\caption{\footnotesize Logarithmic plot of density $\rho$ vs.  forming pressure $\sigma_1$
obtained through uniaxial strain tests (see the inset on the left) on samples (shown in the inset on the right) with water contents $w=5.5\%$ (circle) and $w=7.5\%$ (triangle). Each curve corresponds to
an average value taken over five tests, deviation was found to be negligible.}
\lb{edom}
  \end{center}
\end{figure}
%%%%%%%%%%%%%%%%%%%%%%%%%%%%%%%%%%%%%%%%%%%%%%%%%%%%%%%%%%%%%%%%%%%%%%

Experimental results relative to the 
evolution of the increment in the void ratio due to plastic strain $\Delta e^p$ (in other words, the 
increment in the void ratio remaining after unloading)
as a function of the yield strength under isotropic compression $p_c$
can be interpreted 
employing the micro-mechanical model proposed by Cooper and Eaton (1962), defined as a the double-exponential law
\beq
\label{firstlaw}
-\frac{\Delta e^p}{e_0}=a_1\,\textup{exp}\left( -\frac{\Lambda_1}{p_c}\right)+a_2\,\textup{exp}\left( -\frac{\Lambda_2}{p_c}\right) ,
\eeq
where (neglecting the elastic increment) $\Delta e^p=e-e_0$, being $e_0$ the initial void ratio and $a_1,\,a_2,\, \Lambda_1,\, \Lambda_2 $
material parameters. The void index $e$ can be computed from the
density $\rho$ (at different values of $\sigma_1$, Fig. \ref{edom}) as
\beq
e=\frac{\rho_t}{\rho(1-w)}-1,
\eeq
so that the evolution law for the plastic increment of the void ratio $\Delta e^p$, eq. (\ref{firstlaw}),
can be calibrated to describe results from uniaxial strain experiments, in which the permanent increment of the void ratio has been measured
 at different forming pressure $p_c=\sigma_1$. In particular, the values reported in Table \ref{first} provide the excellent
 fitting\footnote{The best-fit algorithm \lq FindFit' (available in Mathematica$^{\copyright}$) producing least-squares fits, has been used.
The algorithm is based on the minimization of the quantity $\chi^2=\Sigma_i|r_i|^2$, where $r_i$ are residuals giving the difference
between each original data point and its fitted value.} of the evolution law (\ref{firstlaw}) with the experimental data shown in Fig. \ref{1law}.
Note that the difference (visible in Fig. \ref{edom}) of approximately 8\% in the density between the results pertaining to the two water contents  
is also present, though hardly visible, 
in the plots reported in Fig. \ref{1law}.

\begin{table}[!htcb] \caption{\footnotesize{Evolution parameters defining the law of density growth with forming pressure}}
\label{first}
\begin{center}
\begin{tabular}{lll}
\toprule
\textbf{w}  & \textbf{5.5\%}  &  \textbf{7.5\%} \\
\midrule
$a_1$ & $0.78$ & $0.73$ \\
$a_2$ &$0.09$ & $0.14$ \\
$\Lambda_1$ [MPa] & $1.14$ & $0.55$ \\
$\Lambda_2$ [MPa] & $40.86$ & $17.26$ \\
\bottomrule
\end{tabular}
\end{center}
\end{table}

%%%%%%%%%%%%%%%%%%%%%%%%%%%%%%%%%%%%%%%%%%%%%%%%%%%%%%%%%%%%%%%%%%%%%%
\begin{figure}[!htcb]
  \begin{center}
      \includegraphics[width= 14 cm]{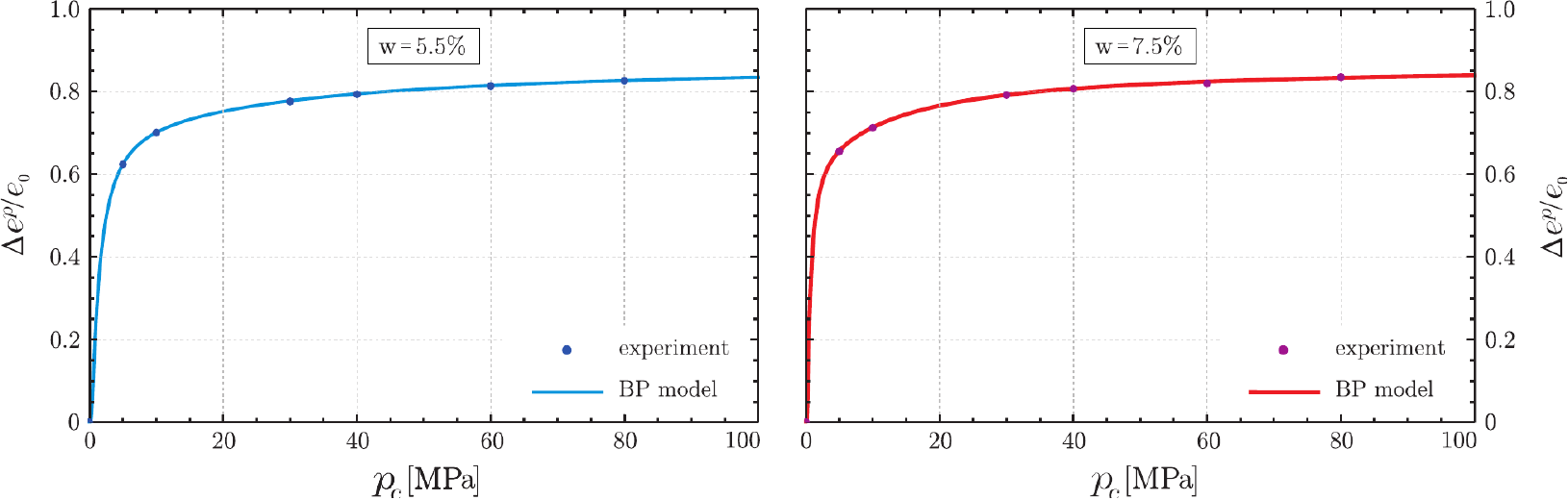}
\caption{\footnotesize Plastic increment in the void ratio $\Delta e^p$ as a
function of the forming pressure $p_c$. Experimental data (dots) are compared with the 
Cooper and Eaton model, eq. (\ref{firstlaw}), (solid line) for two water contents: 
$w=5.5\%$ (left) and $w=7.5\%$ (right). Note that there is a $\approx$8\% (hardly visible) difference 
between the experimental values reported in the two graphs, due to the different water content. 
}
\lb{1law}
  \end{center}
\end{figure}
%%%%%%%%%%%%%%%%%%%%%%%%%%%%%%%%%%%%%%%%%%%%%%%%%%%%%%%%%%%%%%%%%%%%%%

\section{The yield locus and the failure envelope for aluminum silicate green bodies}\lb{calib}
\setcounter{equation}{0}

The aim of this section is the determination of the yield locus and the failure envelope for a green body (made up of aluminum silicate), formed at a specific pressure, through equi-biaxial flexure tests, and high-pressure triaxial compression
and extension tests. The forming pressure has been selected as $p_c^{*}=$40 MPa, a representative value for the stress
states of green bodies used in the industrial production of ceramic tiles.
Eventually, it is shown that the experimental yield locus is tightly fitted by the yield surface proposed by Bigoni and Piccolroaz (2004), while the failure envelope results to be a curved Coulomb criterion.

\subsection{Equi-biaxial flexure test}\label{daidaidai}

With the purpose of obtaining the uniaxial tensile strength, equi-biaxial flexure tests have been performed
on tablets, formed at uniaxial pressure $\sigma_1=40$ MPa with the technique described in Sect. \ref{ciccia}, following the
 ASTM C 1499-05 \lq Standard Test Method  for Monotonic Equi-biaxial Flexural Strength of Advanced Ceramics at Ambient Temperature'.
 The  ASTM standards provide the equi-biaxial flexural strength $\sigma_f$ from the maximum-load $F$ during the test carried until failure as
\beq
\lb{normat}
\sigma_f=\frac{3F}{2 \pi h^2}\left[(1-\nu)\frac{D_S^2-D_L^2}{2D^2}+(1+\nu)\textup{ln}\frac{D_S}{D_L}\right],
\eeq
where $h$ is the height of the green body, $D$, $D_S$ and $D_L$ are the diameters of the sample, of the support ring and of the load ring, respectively.
From the elastic phase observed during triaxial compression tests (Sect. \ref{strizza}),
the Poisson's ratio $\nu$ has been evaluated to be equal to 0.35.\footnote{
The value $\nu=0.35$ has been measured during the initial phase of elastic behaviour with a
circumferential extensometer, which unfortunately has been found useless for measuring the behaviour at large plastic deformation.
However, according to eq. (\ref{normat}), the strength of a tablet is weakly dependent on the Poisson's ratio
(as a matter of fact, an error of $25 \%$ in $\nu$ results in an error of only $2 \% $ in the estimated failure stress, $\sigma_f$).
}
With reference to the tests performed on tablets formed with pressure $\sigma_1=40$ MPa,
the biaxial flexure strength has been evaluated to be $\sigma_f=\left\{1.49;1.61\right\}$ MPa for $w=\left\{5.5;7.5\right\}\%$.

\subsection{High-pressure triaxial compression and extension tests}

\subsubsection{Forming of cylindrical samples}

Cylindrical green body samples of about 70 mm final height (see the inset in Fig. \ref{mould}) have been realized to perform high-pressure triaxial tests.
The samples have been formed through 8 successive layers, each layer has been obtained by filling a (38.2 mm diameter)
mould with 20 g of powder and applying a vertical compression $\sigma_1=3$ MPa. After the eight layers have been realized, a final compression of $\sigma_1=40$ MPa has been applied.
Instead of a single action device, a double action compression device
(where the mould is supported by external springs, see the insets in Fig. \ref{mould}, left) has been used
in order to reduce the inhomogeneity in the density along the height introduced by the friction between the mould and the material.

To check the density distribution along the sample height, two green body samples for each water content have been produced, one with the single action and one with the double action device. 
From the obtained samples, six thick disks have been cut (with a saw) and the density of each disk 
has been measured with a powder pycnometer (using natural marine sand with a mean diameter of 100 $\mu$m,
dried through exposition at 100 $^\circ$C for one hour, with the procedure explained by Della Volpe et al., 2007).
The distribution of the envelope density along the height of the cylindrical samples is shown in Fig. \ref{mould} for green bodies obtained through
double action and single action compression device. Note that 
the density is more uniform for double action than for single action forming, so that in the latter case the density in the lower part
of the cylinder is lower than in the upper part.
%%%%%%%%%%%%%%%%%%%%%%%%%%%%%%%%%%%%%%%%%%%%%%%%%%%%%%%%%%%%%%%%%%%%%%
\begin{figure}[!htcb]
  \begin{center}
      \includegraphics[width= 15 cm]{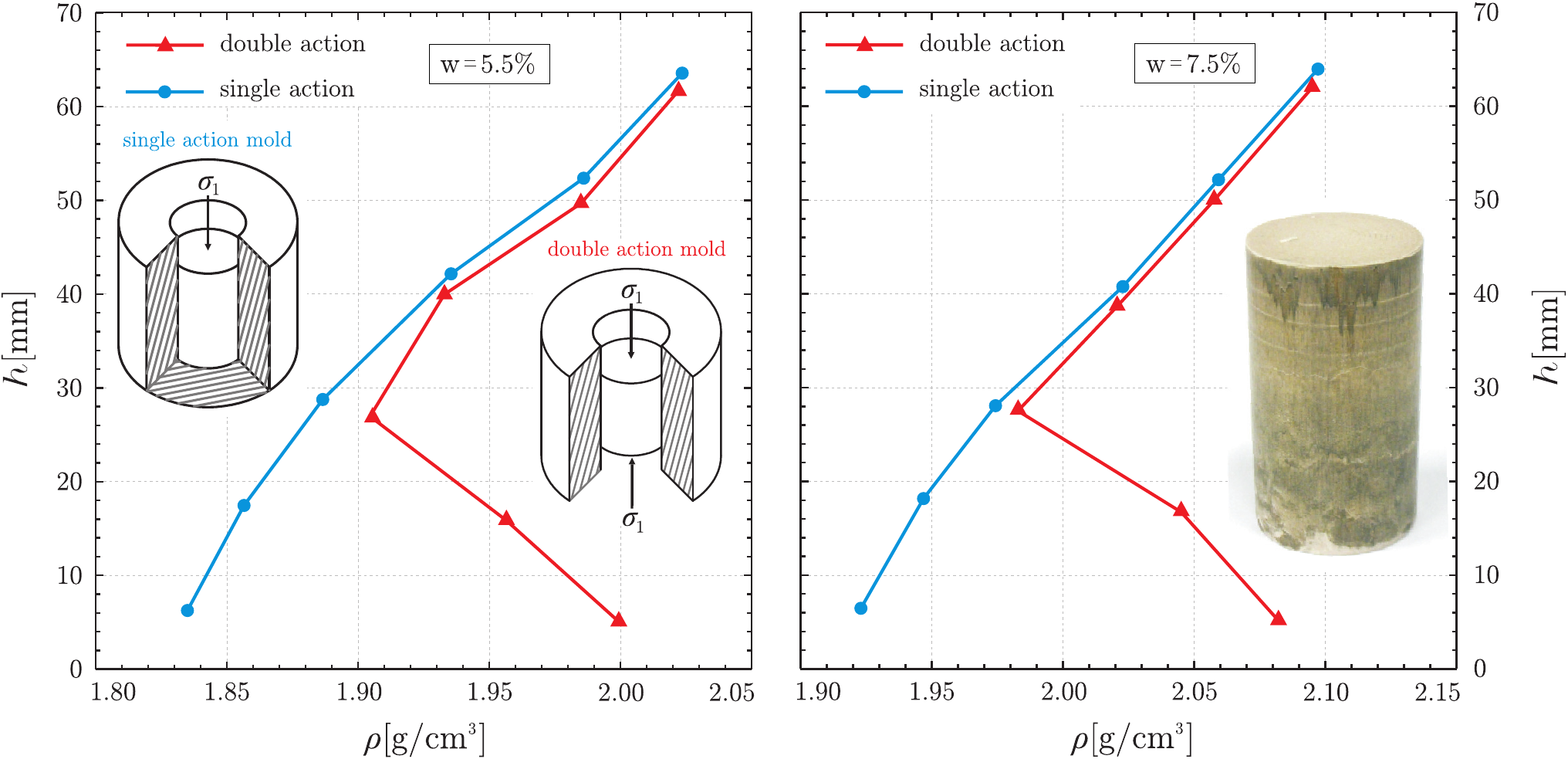}
\caption{\footnotesize Mean (across the transversal area) density $\rho$
distribution along the sample height $h$ for water contents $w=5.5\%$ (left) and $7.5\%$ (right).
For each water content, the double pressure forming
technique with a final pressure $\sigma_1=40$ MPa is compared to the that obtained with the single action device.
Note that the distributions of density are significatively more uniform in the former case than in the latter.}
\lb{mould}
  \end{center}
\end{figure}
%%%%%%%%%%%%%%%%%%%%%%%%%%%%%%%%%%%%%%%%%%%%%%%%%%%%%%%%%%%%%%%%%%%%%%
The deviations from uniformity with the double action device have been found limited to 5\%, which can be considered negligible
for samples to be investigated through triaxial tests.

\subsubsection{Testing protocol and stress-strain curves}\label{strizza}

High-pressure triaxial compression and extension tests have been performed,  according to the ASTM D 2664 95a \lq Standard
Test Method for Triaxial Compressive Strength of Undrained Rock Core Specimens Without Pore Pressure Measurements',
on cylindrical specimens ($38.2$ mm diameter and $\approx 70$ mm height), pre-compacted at $\sigma_1=40$ MPa,
to reproduce the stress conditions during ordinary tile production.

The tests have been performed under vertical displacement control with a ZwickRoell Z250 electromechanical testing machine.
The samples have been inserted into a 45-D0554 Hoek cell, equipped with a rubber sealing sleeve 45-D0554/1 to separate
the specimen from the cell fluid (both manufactured by Controls).  The high lateral pressure has been given by a MAXIMATOR\textsuperscript{\textregistered}  power pack for oil (pressure ratio 1:115, maximum outlet pressure: 1000 bar, maximum air drive pressure: 11 bar, aluminium tank of 6.5 liter) in which the pressure is generated by means of a pneumatically operated pump connected to the compressed-air duckwork system. This power pack is connected to a Parker SensoControl\textsuperscript{\textregistered} SCM 300 through a Parker PD-6000 pressure transducer (0-600 bar).
The data have been acquired with a NI CompactDAQ system, interfaced with Labview SignalExpress (National Instruments).
The testing set-up is shown in Fig. \ref{strum}.

%%%%%%%%%%%%%%%%%%%%%%%%%%%%%%%%%%%%%%%%%%%%%%%%%%%%%%%%%%%%%%%%%%%%%%
\begin{figure}[!htcb]
  \begin{center}
      \includegraphics[width= 11 cm]{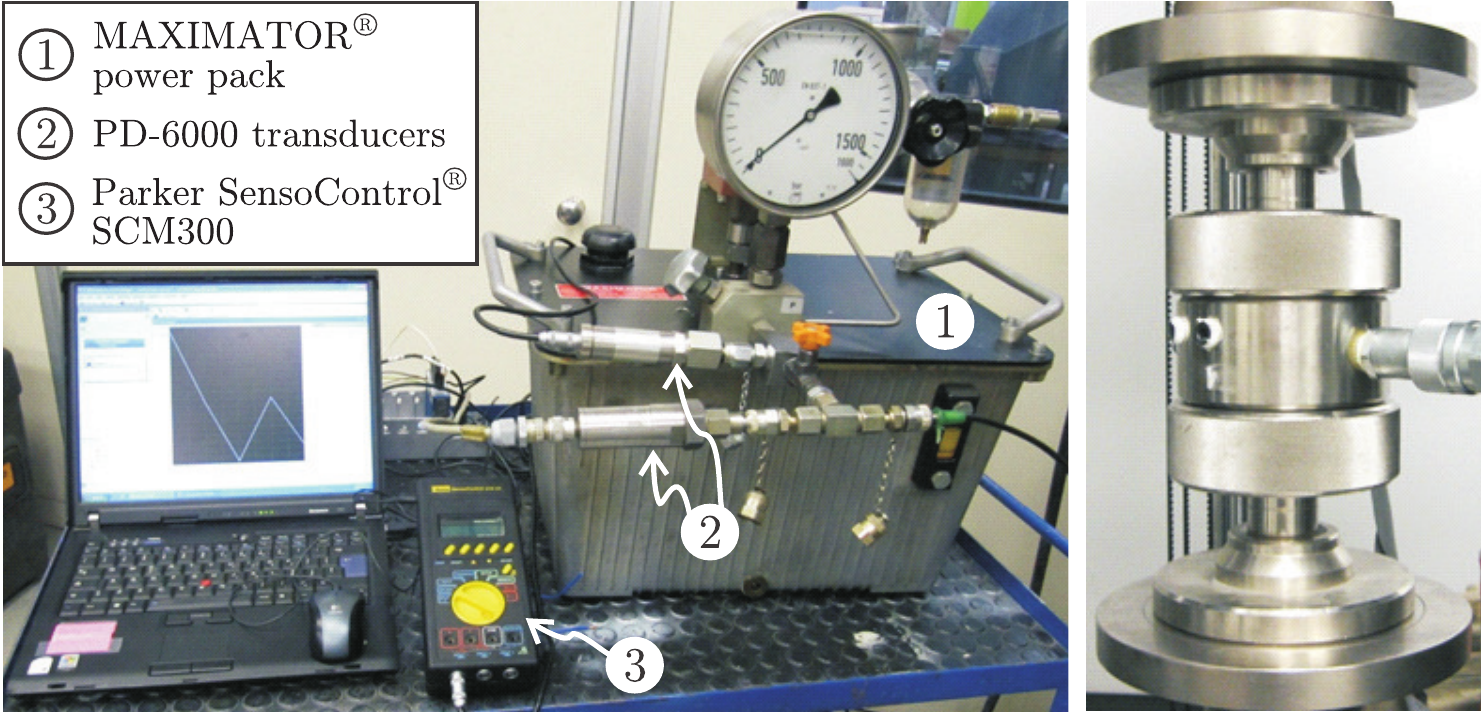}
\caption{\footnotesize  The experimental set-up for the high-pressure triaxial loading tests. Note the equipment
to measure the lateral pressure (on the left) and the 45-D0554 Hoek cell inserted in the ZwickRoell Z250 electromechanical testing machine (on the right).}
\lb{strum}
  \end{center}
\end{figure}
%%%%%%%%%%%%%%%%%%%%%%%%%%%%%%%%%%%%%%%%%%%%%%%%%%%%%%%%%%%%%%%%%%%%%%

Different confining pressures are employed for the investigation of the yield locus for the green body formed at 40 MPa.
The yield locus has been identified by imposing the
coefficient of determination of the line that interpolates the initial points (in the elastic range) as $R^2=0.95$;
the compression tests have been terminated at the critical state, corresponding to the increase of deformation at vanishing stress rate, while failure has not been reached for the extension tests.
In particular, the following testing protocol has been followed, in which the sample:
\begin{itemize}
\item[i)]  has been isotropically loaded (in compression) up to the forming pressure
$$\sigma_1=\sigma_2=\sigma_3=p_c^*=40~\textup{MPa;}$$
\item[ii)]  has been isotropically unloaded up to the value of pressure required for each test, namely,
$$\sigma_1=\sigma_2=\sigma_3=\left\{2,\,5,\,10,\,15,\,20,\,30\right\}~\textup{MPa;}$$

\item[iii.a)] \textit{for triaxial compression}:
has been loaded by increasing the axial stress $\sigma_1$ up to the critical state, keeping constant the cell pressure $\sigma_2=\sigma_3$;

\item[iii.b)] \textit{for triaxial extension}:
has been loaded by increasing the cell pressure $\sigma_2=\sigma_3$ up to a stress beyond the yield point (but not up to failure), keeping constant the axial load $\sigma_1$.
\end{itemize}

Several attempts have been done to measure the radial and circumferential strain (in the transversal section of the sample) during the test, but eventually
it has been impossible to either attach strain gauges to the surface of the green body or to use a circumferential
extensometer (in the range of large plastic deformation). Therefore, the variation of the transversal area of the sample has been eventually estimated by assuming volumetric incompressibility.

The deviatoric stress $q=\left|\sigma_1-\sigma_3\right|$ as a function of the axial deformation $\epsilon_1$
measured during triaxial compression and extension tests is reported in Fig. \ref{triax} and \ref{extension}, respectively.
Note that 
the curves are the average values from three different tests, performed in nominally identical conditions. The maximum discrepancy between the $q-p$ values measured at yielding for the nominally identical tests has been found to be confined within 33\% for w=5.5\% and 17\% for w=7.5\%.
%%%%%%%%%%%%%%%%%%%%%%%%%%%%%%%%%%%%%%%%%%%%%%%%%%%%%%%%%%%%%%%%%%%%%%
\begin{figure}[!htcb]
  \begin{center}
      \includegraphics[width= 15 cm]{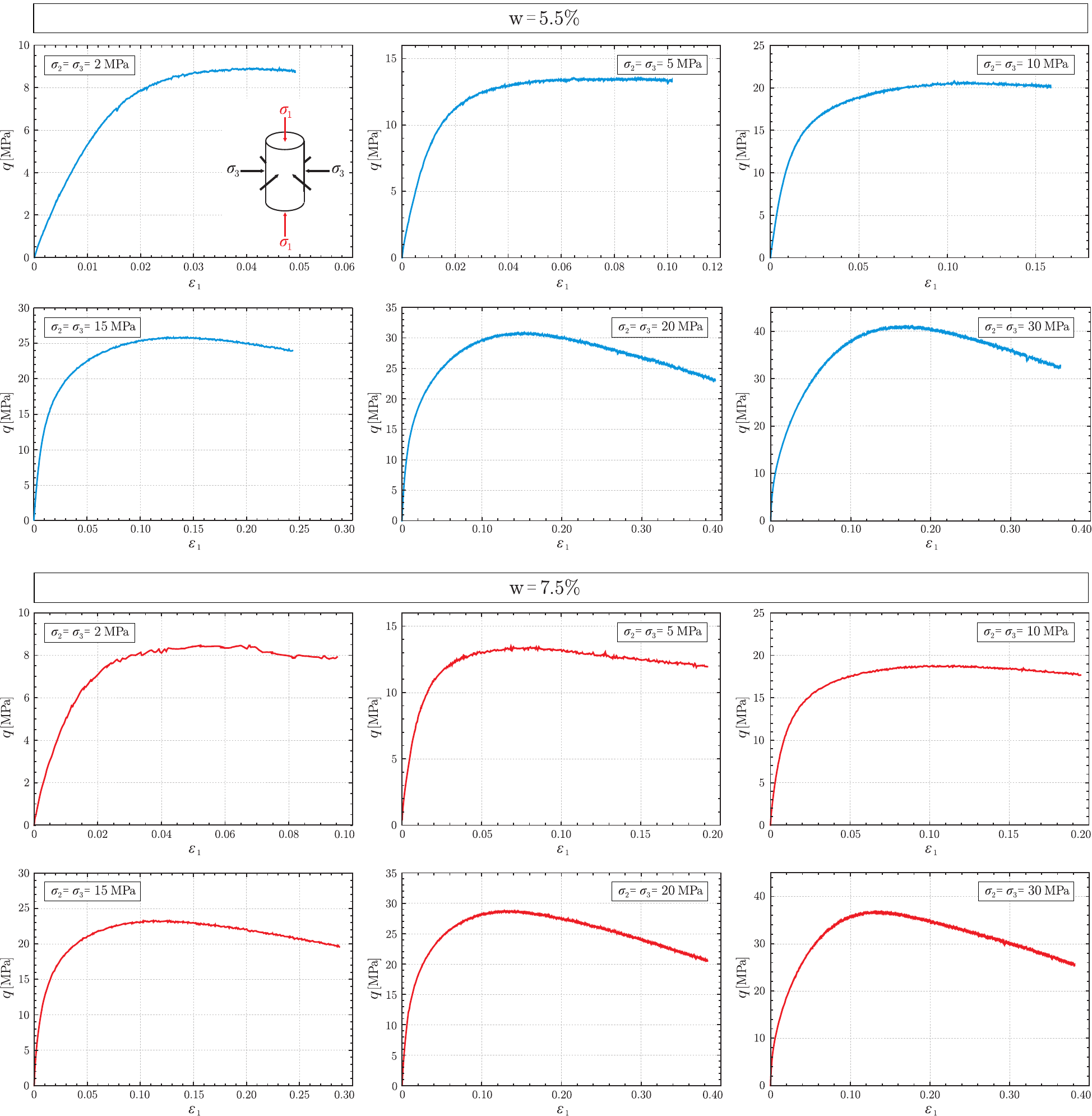}
\caption{\footnotesize Triaxial compression tests at different cell pressures $\sigma_2=\sigma_3= \{$2, 5, 10, 15, 20, 30$\}$ MPa (from left to right) on samples with $w=5.5\%$ (upper part) and 
$w=7.5\%$ (lower part) 
of water content. Each curve represents the average of values from three different tests with the same set-up
(water content and lateral pressure).}
\lb{triax}
  \end{center}
\end{figure}
%%%%%%%%%%%%%%%%%%%%%%%%%%%%%%%%%%%%%%%%%%%%%%%%%%%%%%%%%%%%%%%%%%%%%%
%%%%%%%%%%%%%%%%%%%%%%%%%%%%%%%%%%%%%%%%%%%%%%%%%%%%%%%%%%%%%%%%%%%%%%
\begin{figure}[!htcb]
  \begin{center}
      \includegraphics[width= 15 cm]{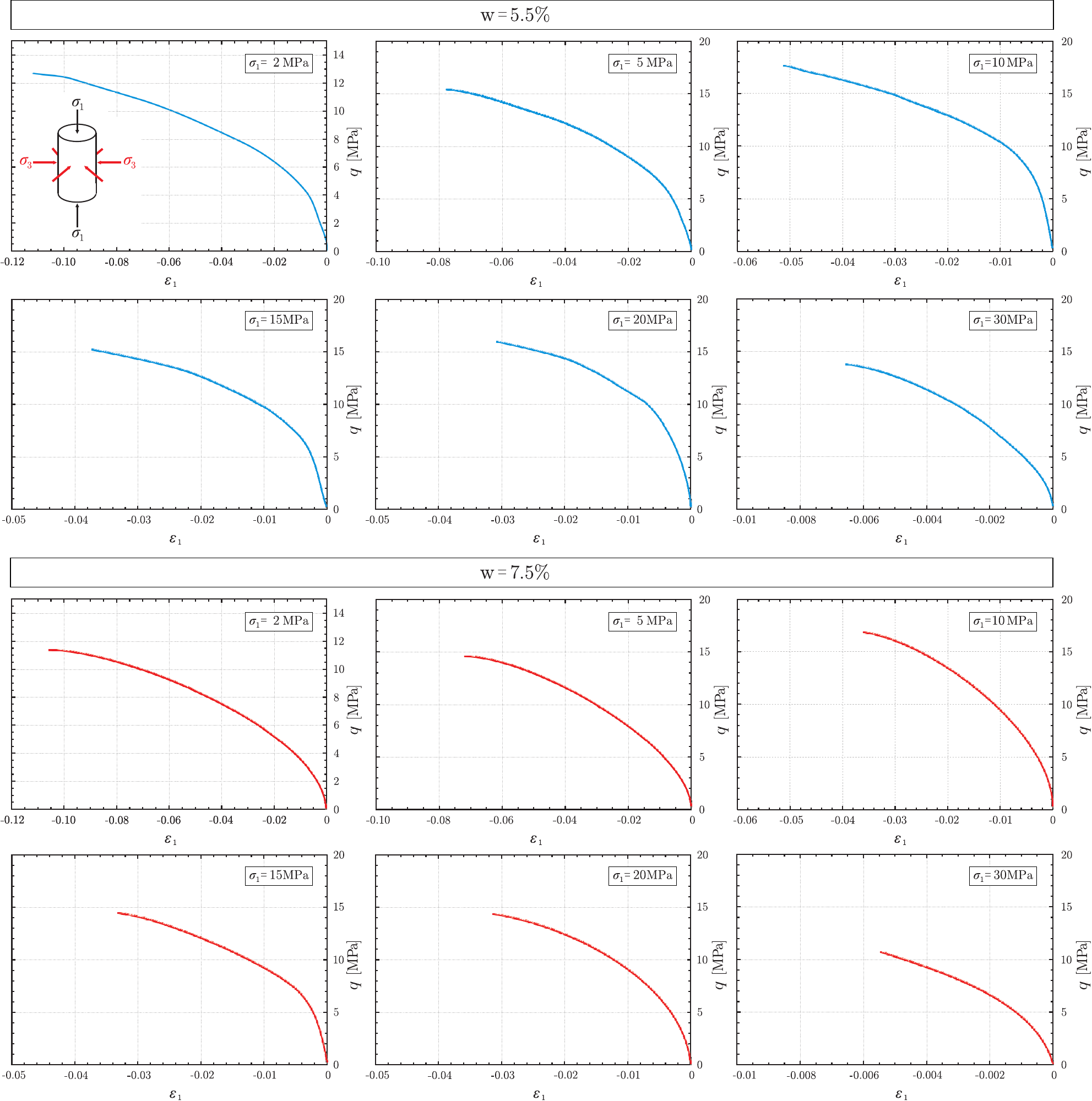}
\caption{\footnotesize Triaxial extension tests at different axial stress $\sigma_1 = \{$2, 5, 10, 15, 20, 30$\}$ MPa (from left to right) on samples with $w=5.5\%$ (upper part) and 
$w=7.5\%$ (lower part) 
of water content. Each curve represents the average of values from three different tests with the same set-up
(water content and axial stress). The tests have been terminated before failure, which would have required a cell pressure higher than the maximum available.}
\lb{extension}
  \end{center}
\end{figure}
%%%%%%%%%%%%%%%%%%%%%%%%%%%%%%%%%%%%%%%%%%%%%%%%%%%%%%%%%%%%%%%%%%%%%%
It can be noted from the figures that the yield stress invariant $q$ initially increases with the confining pressure $p$ and eventually decreases, while the failure stress (reached only for compression tests, Fig. \ref{triax}) always increases with the confining pressure.

\subsection{Meridian sections of the yield locus for triaxial compression and extension}

The yield points measured during triaxial tests at different cell pressures can be reported as functions
of the mean pressure $p=\left(\sigma_1+2\sigma_3\right)/3$, thus defining two meridian sections of the yield locus
(corresponding to Lode's angles $\theta=0$ and $\pi/3$ for triaxial extension and compression tests, respectively)
in the $q-p$ plane. These yield points are complemented
by: (i.) the value of the forming pressure (providing the yield point at $q=0$ 
and $p=p_c^*$) and (ii.) 
the result from the equi-biaxial test,
corresponding, for $\theta=0$, to the yield point 
of coordinates 
$p=-\sigma_f/3$ and $q=\sigma_f$
 and providing
the only available datum on the tensile zone in the $q-p$ representation.

The experimental data in the $q-p$ plane can be interpreted with the BP
yield surface (Bigoni and Piccolroaz, 2004; Bigoni, 2012), that, for triaxial compression and extension, can be written as
\beq
\lb{snervmod}
q=M p_c^{*} \sqrt{\left(\phi-\phi^m\right)\left[2(1-\alpha)\phi+\alpha\right]}
\times \left\{ \barr{ll}
g(\pi/3), & \qquad\textup{triaxial compression},\\
\\
g(0), & \qquad\textup{triaxial extension},
\earr
\right.
\eeq
where $M$ is a parameter defining the pressure-sensitivity of yielding,
$p_c^{*}$ is the yield pressure obtained in an isotropic compression test (assumed equal to the forming pressure $p_c^{*}=40$ MPa)
\beq\label{cazzo}
\phi=\frac{p+c^{*}}{p_c^{*}+c^{*}},
\eeq
$c^{*}$ is the yield strength for isotropic tension and
\beq\label{mazzo}
\left. \barr{lll}
g(0)\\
\\
g(\pi/3)
\earr
\right\} = \frac{1}{\cos\left[\beta\frac{\pi}{6}-\frac{1}{3}\cos^{-1}\left(\pm\gamma\right)\right]},
\eeq
where the following restrictions on the material parameters are necessary to preserve convexity
\beq
M>0,   \,\,\,\,\,   c^*\ge0,   \,\,\,\,\, 0<\alpha<2,  \,\,\,\,\,   m>1, \qquad
0\le\beta\le2,\,\,\,\,\, 0\le\gamma<1.
\eeq

Fixing $\gamma=0.9$,\footnote{This value provides a deviatoric section fairly close to a piecewise linear deviatoric section.}
the remaining five parameters $\left\{\alpha,\,\beta,\, m,\,M, \,c^*\right\}$ have been identified with the values reported in Table \ref{meridian}, 
through an application of the same best-fit algorithm employed in Sect. \ref{ciccia}.
%%%%%%%%%%%%%%%%%%%%%%%%%%%%%%%%%%%%%%%%%%%%%%%%%%%%%%%%%%%%%%%%%%%%%%
\begin{table}[!htcb]
\caption{\footnotesize{Values of the material parameters defining the BP yield locus for a green body of aluminum silicate formed under uniaxial strain at $\sigma_1=$40 MPa
and with water contents of 5.5\% and 7.5\% }} 
\label{meridian}
\begin{center}
\begin{tabular}{lll}
\toprule
\textbf{w}  & \textbf{5.5\%}  &  \textbf{7.5\%} \\
\midrule
$\alpha$ & $1.95$ & $1.95$ \\
$m$ & $4.38$ & $4.57$ \\
$M$ & $0.25$ & $0.22$ \\
$\beta$ & $0.10$ & $0.08$ \\
$\gamma$ & $0.90$ & $0.90$ \\
$c^{*}$ [MPa] & $0.94$ & $1.23$ \\
\bottomrule
\end{tabular}
\end{center}
\end{table}
%%%%%%%%%%%%%%%%%%%%%%%%%%%%%%%%%%%%%%%%%%%%%%%%%%%%%%%%%%%%%%%%%%%%%%

The meridian sections 
(for Lode's angles $\theta=\pi/3$ and $\theta=0$) 
of the yield locus, corresponding to the parameters listed in Table \ref{meridian}, 
are reported in the $q-p$ plane in Fig. \ref{meridiana}, together with experimental points.
%%%%%%%%%%%%%%%%%%%%%%%%%%%%%%%%%%%%%%%%%%%%%%%%%%%%%%%%%%%%%%%%%%%%%%
\begin{figure}[!htcb]
  \begin{center}
      \includegraphics[width= 14 cm]{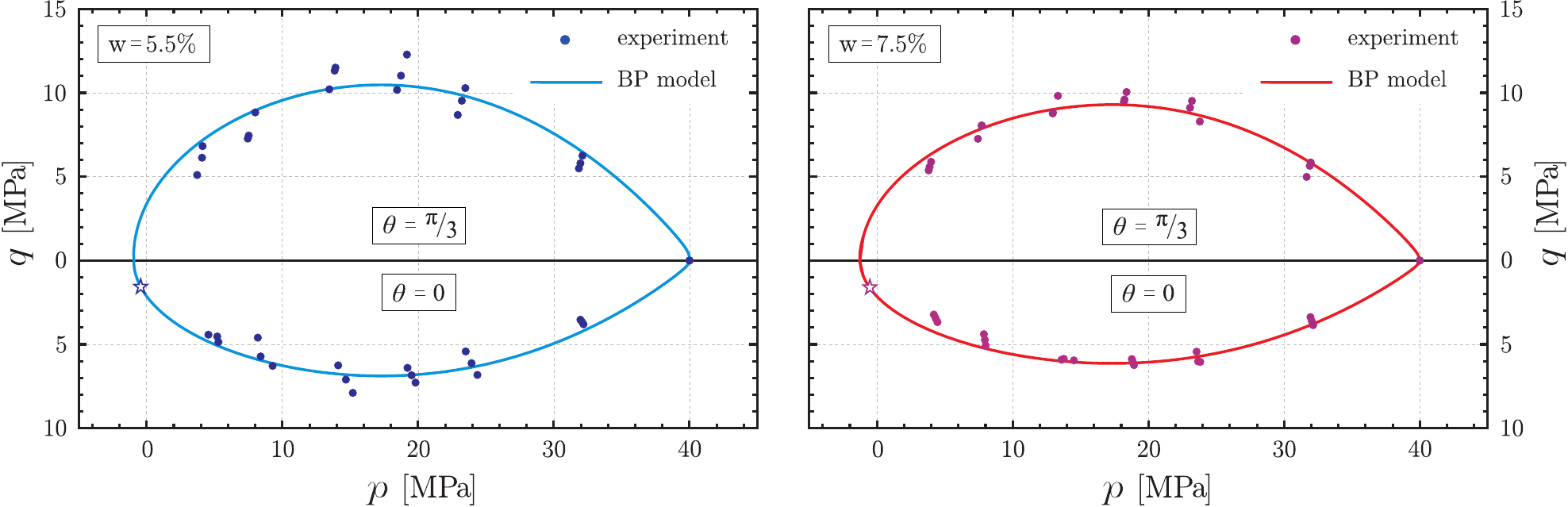}
\caption{\footnotesize Meridian sections of the yield locus at Lode's angles
$\theta=\pi/3$ and $\theta=0$ for aluminum silicate green body formed under uniaxial
strain at $p_c^*=\sigma_1=40$ MPa, with water contents of 5.5\% (left) and 7.5\% (right). Yield points from triaxial tests (dots),
evaluated through linear regression,
and from equi-biaxial flexure tests (stars) are fitted with the BP yield surface (Bigoni and Piccolroaz, 2004), corresponding to the material parameters
listed in Table \ref{meridian}.}
\lb{meridiana}
  \end{center}
\end{figure}
%%%%%%%%%%%%%%%%%%%%%%%%%%%%%%%%%%%%%%%%%%%%%%%%%%%%%%%%%%%%%%%%%%%%%%

\subsection{The failure envelope}

Defining the \lq critical state' as the $q-p$ values corresponding to the peak of the curves shown in Fig. \ref{triax}, the critical line reported in Fig. \ref{critica} (upper part) has been found. 
From the Mohr representation (lower part of Fig. \ref{critica}),
a mean value of the friction angle equal to 24$^\circ$ 
for w=5.5\% (22.5$^\circ$ for w=7.5\%) 
can be deduced.
%%%%%%%%%%%%%%%%%%%%%%%%%%%%%%%%%%%%%%%%%%%%%%%%%%%%%%%%%%%%%%%%%%%%%%
\begin{figure}[!htcb]
  \begin{center}
      \includegraphics[width= 14 cm]{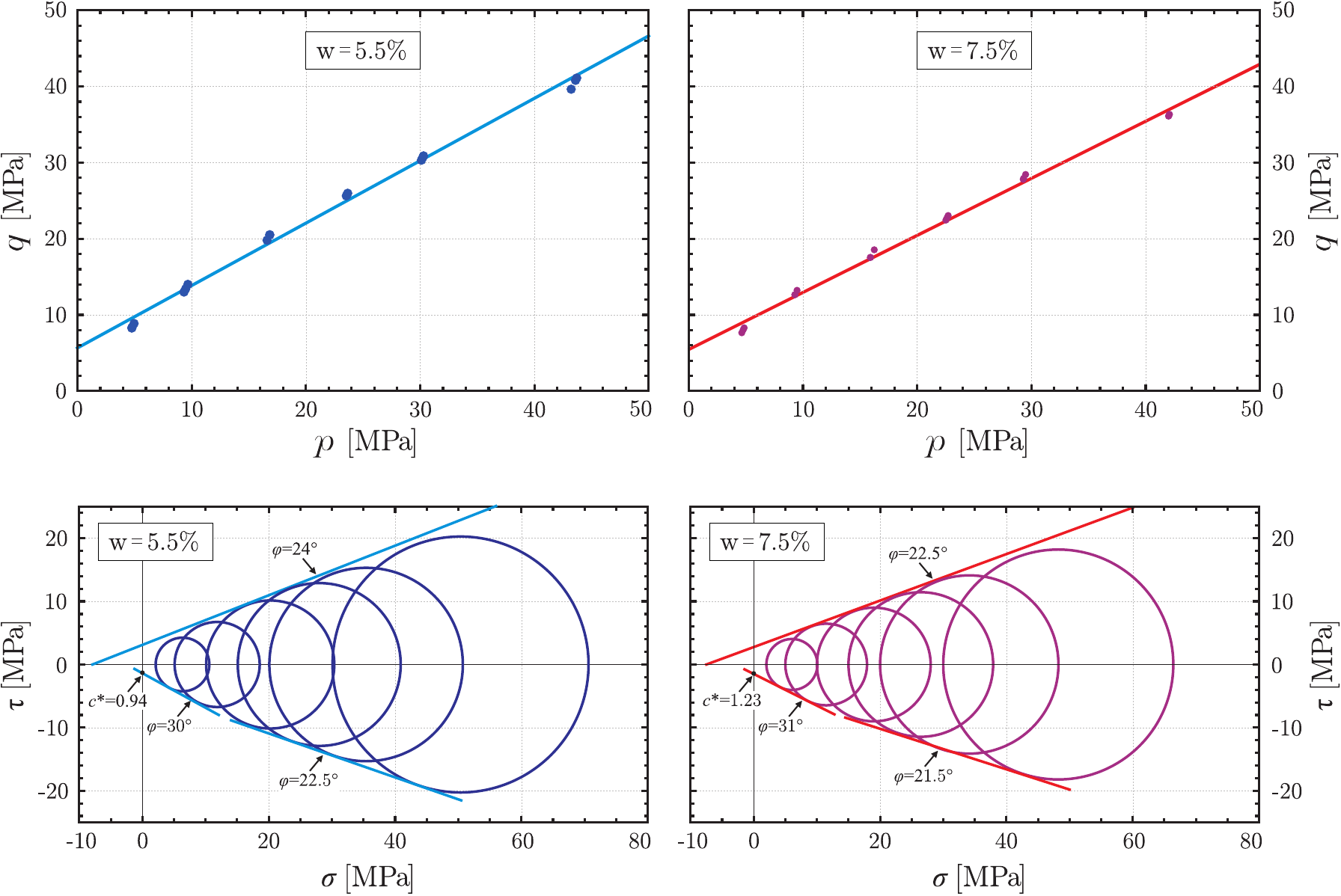}
\caption{\footnotesize Upper part: critical line (corresponding to the peaks of the curves shown in Fig. \ref{triax}), for aluminum silicate green body formed under uniaxial
strain at 40 MPa and with water contents of 5.5\% (left) and 7.5\% (right). Lower part: Mohr representation of the failure stresses with Coulomb failure envelope; this corresponds to a mean friction angle $\varphi$ of 24$^\circ$ (left) and 22.5$^\circ$ (right). The failure envelope results to be curved, so that higher friction angles occur at small confining pressure.}
\lb{critica}
  \end{center}
\end{figure}
%%%%%%%%%%%%%%%%%%%%%%%%%%%%%%%%%%%%%%%%%%%%%%%%%%%%%%%%%%%%%%%%%%%%%%

Since three nominally identical tests have been performed for every initial $p$, three experimental points are reported in the graphs of the upper part of Fig. \ref{critica}, while the Mohr circles reported in the lower part of the figure have been drawn with the mean values calculated over the three different tests. 
Note that the failure envelope found in the Mohr plane does not correspond to the two Coulomb lines, but is curved with a tangent more inclined at low mean stress than at high. In particular, the friction angle ranges between 30$^\circ$ 
and 22.5$^\circ$ for w=5.5\% (31$^\circ$ and 21.5$^\circ$ for w=7.5\%). Finally, note that the cohesion found in the Mohr plane using the tangent to the two Mohr circles at low pressure coincide with that deduced through the best fitting with the BP yield surface (value of $c^*$ reported in Table \ref{meridian}).

\section{Evolution of cohesion for aluminum silicate green bodies}

A relationship between the cohesion $c$ and the forming pressure $p_c$ can be defined, providing the dependence
of this parameter on the plastic strain. To this purpose, in addition to the equi-biaxial flexure tests (Sect. \ref{daidaidai}) performed on tablets formed at $\sigma_1=40$ MPa, other tests have been performed on tablets formed at different pressures, namely, $\sigma_1=\left\{5,\,10,\, 30,\,60,\,80\right\}$ MPa.
The measured equi-biaxial flexural strengths $\sigma_f$
at different values of forming pressure $\sigma_1$ are reported in Fig. \ref{biax}.
%%%%%%%%%%%%%%%%%%%%%%%%%%%%%%%%%%%%%%%%%%%%%%%%%%%%%%%%%%%%%%%%%%%%%%
\begin{figure}[!htcb]
  \begin{center}
      \includegraphics[width= 8 cm]{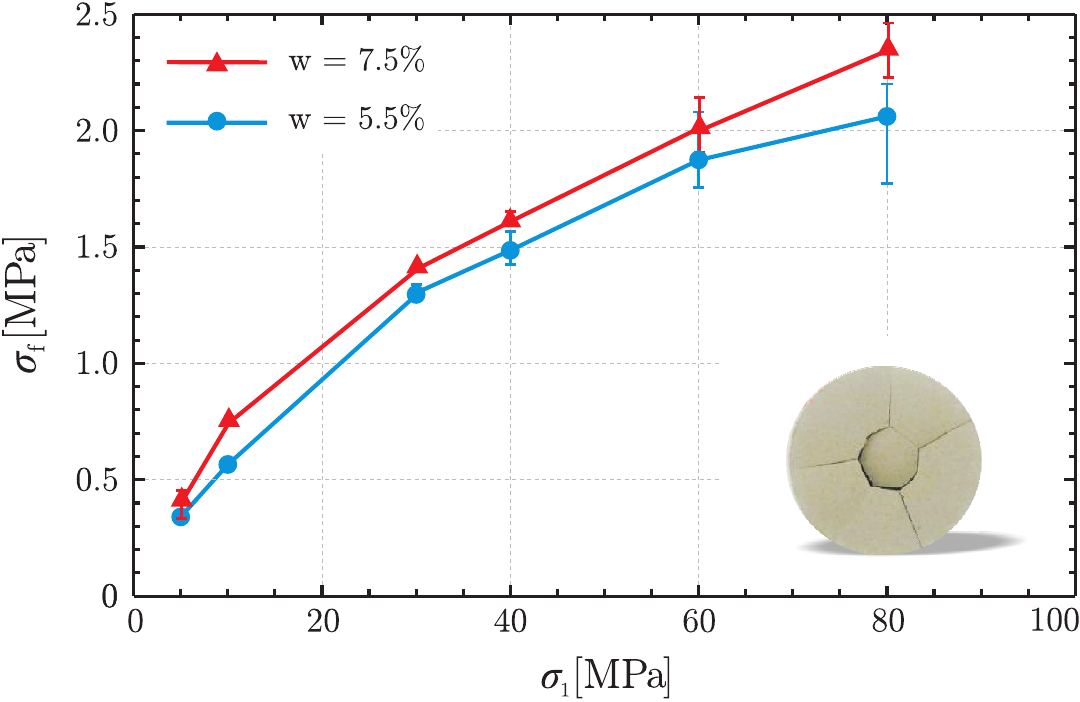}
\caption{\footnotesize Flexure strength $\sigma_f$ vs. forming pressure $\sigma_1$ evaluated from equi-biaxial flexure tests performed on
ceramic tablets with water content $5.5\%$ (circle) and  $7.5\%$ (triangle). The post-mortem sample is shown in the inset.}
\lb{biax}
  \end{center}
\end{figure}
%%%%%%%%%%%%%%%%%%%%%%%%%%%%%%%%%%%%%%%%%%%%%%%%%%%%%%%%%%%%%%%%%%%%%%

The data are interpreted with the model of increase in cohesion proposed by Piccolroaz et al. (2006) in which a limit value of cohesion, $c_{\infty}$, is introduced as 
\beq
\label{secondlaw}
c=c_{\infty} \left[ 1-\textup{exp} \left( -\Gamma \left\langle p_c-p_{cb}\right\rangle\right)\right],
\eeq
where $\left\langle \right\rangle$ is the Macaulay brackets, 
$\Gamma$ is the parameter defining the \lq velocity of growth' of cohesion and $p_{cb}$ is the breakpoint pressure. The values of these
parameters reported in Table \ref{second} 
provide the nice fittings  of
the experimental data shown in Fig. \ref{seclaw} (obtained by means of the same best-fit algorithm adopted in Sect. \ref{ciccia}), thus confirming the validity of the model. 
\begin{table}[!htcb] \caption{Values of the material parameters defining the evolution of cohesion law} 
\label{second}
\begin{center}
\begin{tabular}{lll}
\toprule
\textbf{w}  & \textbf{5.5\%}  &  \textbf{7.5\%} \\
\midrule
$\Gamma$ [MPa$^{-1}$] & $0.06$ & $0.10$  \\
$c_{\infty}$ [MPa] & $1.10$ & $1.35$  \\
$p_{cb}$ [MPa] & $0.22$ & $0.17$ \\
\bottomrule
\end{tabular}
\end{center}
\end{table}

%%%%%%%%%%%%%%%%%%%%%%%%%%%%%%%%%%%%%%%%%%%%%%%%%%%%%%%%%%%%%%%%%%%%%%
\begin{figure}[!htcb]
  \begin{center}
      \includegraphics[width= 14 cm]{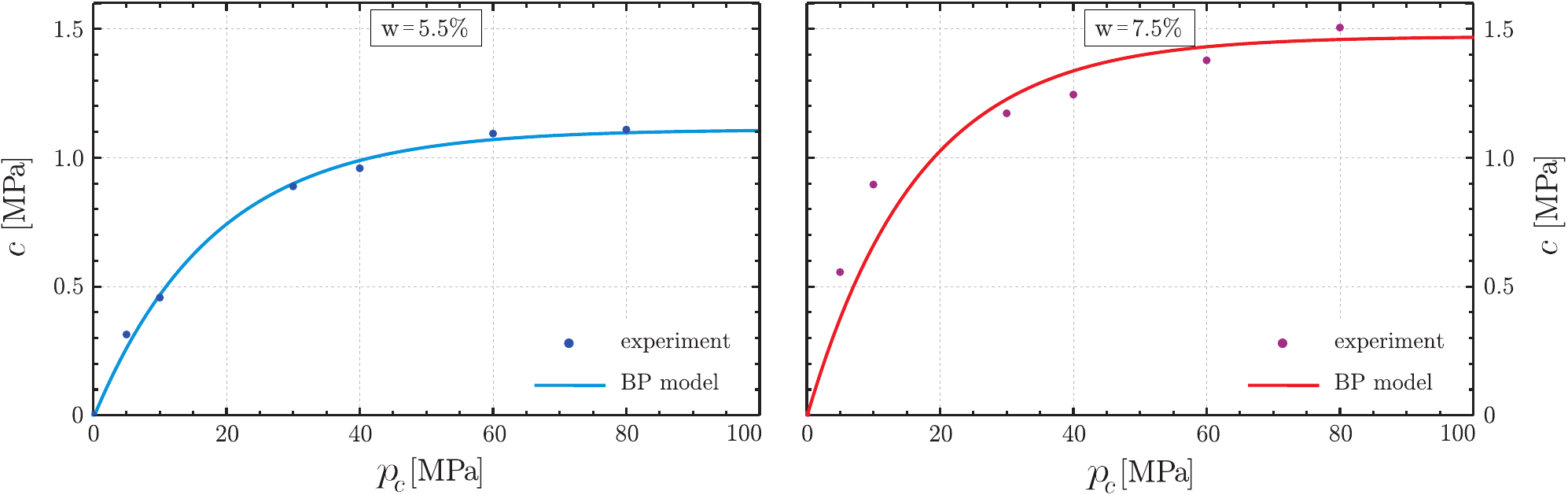}
\caption{\footnotesize Cohesion parameter $c$ as a function of the forming pressure $p_c$.
Experimental data relative to aluminum silicate powder are compared to the model of increase of cohesion,
eq. (\ref{secondlaw}), proposed by Piccolroaz et al. (2006) (solide line), evaluated with the parameters
of Table \ref{second}, for water contents 5.5\% (left) and 7.5\% (right). }
\lb{seclaw}
  \end{center}
\end{figure}
%%%%%%%%%%%%%%%%%%%%%%%%%%%%%%%%%%%%%%%%%%%%%%%%%%%%%%%%%%%%%%%%%%%%%%

%%%%%%%%%%%%%%%%%%%%%%%%%%%%%%%%%%%%%%%%%%%%%%%%%%%%%%%%%%%%%%%%%%%%%%%%%%%%%%%%%%%%%%%%
\section{Conclusions}
Mechanical characterization of the plastic properties has been performed for ceramic green bodies obtained from cold compaction of an 
aluminum silicate powder, usually employed for the industrial production of tiles. The most important
(and previously almost unexplored) feature of the plastic behaviour of the material is (i.) its yield locus,
which has been determined through two meridian sections corresponding to triaxial compression and extension loading paths. Other 
investigated features include: (ii.) the critical state line, determining the stress states corresponding to failure; 
(iii.) the evolution laws of density and (iv.) cohesion with the change of the forming pressure.
The developed testing protocol and the experimental evidence on the plastic behaviour of green bodies 
are fundamental for a proper modelling, simulation and the optimization of the production process
of traditional and advanced ceramics.

%%%%%%%%%%%%%%%%%%%%%%%%%%%%%%%%%%%%%%%%%%%%%%%%%%%%%%%%%%%%%%%%%%%%%%%%%%%%%%%%%%%%%%%%
%\setcounter{equation}{0}
%\appendix
%\section{Appendix}\label{descrizione_esperimenti}

\vspace*{3mm} \noindent {\sl Acknowledgments } Financial support
from the European FP7 - Intercer2 project
(PIAP-GA-2011-286110-INTERCER2) is gratefully acknowledged.

\vspace*{5mm} \noindent

 { \singlespace
}

%%%%%%%%%%%%%%%%%%%%%%%%%%%%%%%%%%%%%%%%%%%%%%%%%%%%%%%%%%%%%%%%%%%%%%%%%%%%%%%%%%%%%

\end{document}